\documentclass{article}
\usepackage{graphicx}

\begin{document}

\title{ Generalized Heisenberg algebra coherent states for Power-law potentials}

\author{{\bf{K. Berrada}}\thanks{berradakamal2006@gmail.com} ,\hspace{0.2cm}\hspace{0.2cm}{\bf{M. El Baz}}\thanks{elbaz@fsr.ac.ma} \hspace{0.2cm} and \hspace{0.2cm} {\bf{Y. Hassouni}}\thanks{y-hassou@fsr.ac.ma} }

\date{}
\vspace{-0.6cm} \maketitle{ \begin{center} \it Laboratoire de
Physique Th\`eorique, D\'epartement de Physique \linebreak Facult\'e
des sciences, Universit\'e Mohammed V - Agdal \linebreak Av. Ibn
Battouta, B.P. 1014, Agdal, Rabat, Morocco \end{center}}

\date{}
\maketitle

\vspace{1cm}

\begin{abstract}
Coherent states for power-law potentials are constructed using
generalized Heisenberg algabras. Klauder's minimal set of conditions
required to obtain coherent states are satisfied. The statistical
properties of these states are investigated through the evaluation
of the Mandel's parameter. It is shown that these coherent states
are useful for describing the states of real and ideal lasers.
\end{abstract}

\section{Introduction}
Ever since the introduction of coherent states theory, several works
have been focused on these states and they have turned out to be a
very useful tool for the investigation of various problems in
physics and have offered surprisingly rich structures
\cite{1,2,3,4}. Coherent states were first introduced by
Schr\"{o}dinger \cite{5} in the context of the harmonic oscillator,
and was interested in finding quantum states which provide a close
behavior to the classical one. Later, the notion of coherent had
become very important in quantum optics thanks to the works of
Glauber \cite{6}. He defined these states as eigenstates of the
annihilation operator $\hat{a}$ of the harmonic oscillator, while
demonstrating that they have the interesting property of minimizing
the Heisenberg uncertainty relation. The notion of coherent states
was extended afterwards to different systems other the harmonic
oscillator for which they were originally built. Coherent states
associated to $SU(2)$ and $SU(1,1)$ algebras were introduced by
Peremolov \cite{7,8}. These states describe several systems and also
have many applications in quantum optics, statistical mechanics,
nuclear physics, and condensed matter physics \cite{9,10,11}.
Generally, there are several definitions of coherent states. The
first is to define the coherent states as eigenstates of an
annihilation group operator with complex eigenvalues. This
definition is called Barut-Girardello coherent states \cite{12}. The
second definition, often called Peremolov coherent states
\cite{7,13}, is to construct the coherent states by acting with an
unitary $z-$displacement operator on the ground state of the system,
with $z\in \mathcal{C}$. The last definition, often called
intelligent coherent states \cite{14,15}, is to consider that the
coherent states give the minimum-uncertainty value $\Delta x\Delta
p={\hbar\over2}$. These three definitions are equivalent only in the
special case of the Heisenberg$-$Weyl group (Harmonic oscillator).
Due to this multitude of definitions of the notion of coherent
states, Klauder \cite{9} proposed a set of conditions to be obeyed
by any set of coherent states. These conditions are in fact the
common properties satisfied by coherent states in all the
definitions presented before. This way this set of condition
constitutes the minimum set of requirements on a set of states in
order to deserve the nomenclature {\it coherent states}.

Generally, it is difficult to build coherent states for arbitrary
quantum mechanical systems. But for systems with discrete spectrums,
there is an elegant method of building coherent states which has
been recently proposed in \cite{1} basing on generalized Heisenberg
algebra. These coherent states are defined as eigenstates of the
annihilation operator of the generalized Heisenberg algebra having
infinite$-$dimensional representations and satisfying the minimum
set of conditions required to construct Klauder's coherent states.

Coherent states for harmonic oscillator and other systems have
attracted much attention for the past years
\cite{16,17,18,19,20,21,22} and play important roles in different
contexts. The power-law potentials (a general class of potentials)
have many applications in theoretical and experimental physics,
which can be used to describe a large class of quantum mechanical
systems \cite{23,24,25,26}. The coherent states of these potentials
could be very helpful and bring more insights on these subjects. In
this letter, we are going to construct the coherent states for these
potentials following the formalism used in \cite{1}. We, then,
investigate the statistical properties of these coherent states. It
will be shown that these states can exhibit a sub-Poissonian,
Poissonian or a super-Poissonian distribution depending on the
parameters of the system.

The organization of the letter is as follows. In section 2 we
describe the construction of generalized Heisenberg algebra coherent
states; in section 3 we build coherent states for power-law
potentials and investigate their statistical properties using
Mandel's $Q$ parameter; in the last section, we present our
conclusion.

\section{Coherent states of generalized Heisenberg algebra}
In this letter, we are going to use the generalized Heisenberg
algebra ($GHA$) which has been introduced in Ref. \cite{27} and
construct the coherent states associated. The version of the $GHA$
we are going to describe is defined by the following commutation
relations
\begin{equation}\label{h1}
HA^\dag=A^\dag f(H),
\end{equation}
\begin{equation}
AH=f(H)A,
\end{equation}
\begin{equation}\label{h3}
\left[A,A^\dag\right]=f(H)-H,
\end{equation}
where its generators $(H, A, A^\dag)$ obey the Hermiticity
properties $A=A^\dag$, $H=H^\dag$. The Casimir element of the $GHA$
is given by
\begin{equation}
C=A^\dag A-H=AA^\dag-f(H).
\end{equation}
Here $H$ is the Hamiltonian of the physical system under
consideration and $f(H)$ is an analytic function of $H$, called the
characteristic function of the algebra. It was shown in Ref.
\cite{27} that a large class of Heisenberg algebras can be obtained
by choosing the appropriate function $f(H)$. For example, by
choosing the characteristic function as $f(H)=H+1$, the algebra
described by Eqs. (\ref{h1})$-$(\ref{h3}) becomes the harmonic
oscillator algebra and for $f(H)=qH+1$ we obtain the deformed
Heisenberg algebra.

The irreducible representation of the $GHA$ is given through a
general vector $|m\rangle$ which is required to be an eigenstate of
the Hamiltonian. It is described as follows
\begin{equation}
H|m\rangle=E_m|m\rangle
\end{equation}
\begin{equation}
A^\dag|m\rangle=N_m|m+1\rangle
\end{equation}
\begin{equation}
A|m\rangle=N_{m-1}|m-1\rangle
\end{equation}
with
\begin{equation}
N_m^2=E_{m+1}-E_0,
\end{equation}
where $E_m=f^m(E_0)$, the $m^{th}$ iterate of $E_0$ under $f$, are
the eigenvalues of the Hamiltonian $H$. $A^\dag$ and $A$ are the
generalized creation and annihilation operators of the $GHA$. This
$GHA$ describes a class of quantum systems having energy spectra
written as
\begin{equation}
E_{n+1}=f(E_n)
\end{equation}
where $E_{n+1}$ and $E_n$ are successive energy levels.

The coherent states for the $GHA$ were previously investigated in
Ref. \cite{1}. They are defined as the eigenvalues of the annihilation
operator of the $GHA$
\begin{equation}
A|z\rangle=z|z\rangle \qquad\qquad z\in \mathcal{C},
\end{equation}
and can be expressed as
\begin{equation}
|z\rangle=\mathcal{N}(z)\sum_{n=0}^{\infty}{z^n\over
N_{n-1}!}|n\rangle.
\end{equation}
The normalization factor is given by
\begin{equation}
\mathcal{N}(z)=\left(\sum_{n=0}^{\infty}{|z|^{2n}\over
N_{n-1}^2!}|n\rangle\right)^{-{1\over2}},
\end{equation}
where by definition $N_n!=N_0N_1 \cdot\cdot\cdot N_n$ and by
consistency $N_{-1}!=1.$

According to Klauder \cite{9,28}, the minimal set of requirements to be imposed
on a state $|z\rangle$ to be a coherent state is :\\
(i) Normalizability
\begin{equation}
\langle z|z\rangle=1\;;
\end{equation}
(ii) Continuity in the label
\begin{equation}
\left|z-z^{'}\right|\rightarrow0 \qquad \Rightarrow \qquad
\left||z\rangle-|z^{'}\rangle\right|\rightarrow0\;;
\end{equation}
(iii) Resolution of unity
\begin{equation}
\int\int_\mathcal{C}d^2z|z\rangle\langle z|W(|z|^2)=1.
\end{equation}
In general, the first two conditions are easily satisfied. This is
not the case for the third. In fact, this condition restricts
considerably the choice of the states to be considered.

Now, we are going  to analyze the above minimal set of conditions to
obtain a Klauder's coherent states for power-law potentials.

\section{Coherent states for Power-law potentials}

The general expression of a one-dimensional power-law potential is \cite{23}
\begin{equation}
\hat{V}(x,k)=V_0\left|{x\over a}\right|^k,
\end{equation}
where $V_0$ and $a$ are constants with the dimensions of energy and
length, respectively. Here $k$ is the power-law exponent.

These power-law potentials can be used to describe a large class of
quantum systems by the proper choice of the exponent $k$. For $k>2$
we find tightly binding potentials, $k=2$ we have harmonic
oscillator potentials and  $k<2$ expresses loosely binding
potentials.

The Hamiltonian  corresponding to different potentials can be
written in the form
\begin{equation}
\hat{H}={\hat{p}^2\over 2m}+\hat{V}(k,x),
\end{equation}
where the eigenvalue equations are given by
\begin{equation}
\hat{H}(k)|n\rangle=E_{n,k}|n\rangle \qquad\qquad n\geq0.
\end{equation}
The energies $E_{n,k}$ may be obtained with the help of $WKB$
approximation \cite{24,25,26}
\begin{equation}\label{0}
E_{n,k}=\omega(k)\left(n+{\gamma\over4}\right)^{2k\over k+2},
\end{equation}
where
$$\omega(k)=\left[{\pi\over2a\sqrt{2m}}V_0^{1\over k}{\Gamma\left({1\over k}+
{3\over2}\right)\over\Gamma\left({1\over
k}+1\right)\Gamma\left({3\over2}\right)}\right]^{2k\over k+2}$$ is
the effective frequency. Here $\gamma$ defines the Maslov index
which accounts for boundary effects at the classical turning points.

The above equation (\ref{0}) shows that the energy difference
between adjacent levels, $\Delta E_{n,k}=E_{n,k}-E_{n-1,k}\propto
\left(n+{\gamma\over4}\right)^{k-2\over k+2} $, increases for $k>2$
as  $n$ increases (tightly binding potentials). In contrast, for
$k<2$ it decreases with the increase of $n$ (loosely binding
potentials), while for $k=2$ it does not depend on $n$, so the
energy spectrum is equally spaced.

The question now is to find the characteristic function of the
generalized Heisenberg algebra associated with these spectrums. From
the above expression, we have
\begin{equation}\label{1}
n=\left(E_{n,k}\right)^{2k\over k+2}-{\gamma\over4}
\end{equation}
leading us to the expression
\begin{equation}\label{2}
E_{n+1,k}=\left(n+1+{\gamma\over4}\right)^{k+2\over 2k}.
\end{equation}
The substitution of Eq. (\ref{1}) in Eq. (\ref{2}) allows us to
obtain the characteristic equation
\begin{equation}
E_{n+1,k}=\left(\left(E_{n,k}\right)^{k+2\over 2k}+1\right)^{2k\over
k+2}.
\end{equation}
The characteristic function,then, is
\begin{equation}
f(x,k)=\left(\left(x\right)^{k+2\over
2k}+1\right)^{2k\over k+2}.
\end{equation}

According to the scheme prescribed in the previous section, the
coherent states associated to power-law potentials are given by
\begin{equation}\label{3}
|z,k\rangle=\mathcal{N}\left(|z|^2,k\right)\sum_{n=0}^{\infty}{z^n\over\sqrt{g(n,k)}}|n\rangle
\end{equation}
where
\begin{equation}
g(n,k)=\prod_{i=1}^{n}\left(\left(i+{\gamma\over4}\right)^{{2k\over
k+2}}-\left({\gamma\over4}\right)^{{2k\over k+2}}\right),\qquad
g(0,k)=1,
\end{equation}
and the normalization function
\begin{equation}
\mathcal{N}\left(|z|^2,k\right)=\left(\sum_{n=0}^{\infty}{|z|^{2n}\over
g(n,k)}\right)^{-{1\over2}}.
\end{equation}

Now, we can investigate the overcompleteness (or resolution of unity
operator) of the $GHA$ coherent states for power-law potentials
given by Eq. (\ref{3}). To this end, we assume the existence of a
positive Weight function $W(|z|^2)$ so that an integral over the
complex plane exists and gives the result
\begin{equation}\label{4}
\int\int_\mathcal{C}d^2z|z,k\rangle\langle z,k|W(|z|^2)=1.
\end{equation}
Indeed, substituting Eq. (\ref{3}) into Eq. (\ref{4}), adopting the
polar coordinate representation $z=r\hbox{exp}(i\theta)$ of complex
numbers and using the completeness of the states $|n\rangle$, the
resolution of unity can be expressed by
\begin{equation}\label{5}
\int_0^\infty x^n\overline{W}(x)\;dx=g(n,k), \qquad
 n=0,1,2,\cdot\cdot\cdot
\end{equation}
where $x=r^2$ and $\overline{W}(x)=\pi
W(x)\mathcal{N}^2\left(x,k\right)$.

The positive quantities $g(n,k)$ are the power moments of the
function $\overline{W}(x)$ and the problem stated in Eq. (\ref{5})
is the Stieltjes moment problem which can be solved by means of
Mellin and inverse Mellin transforms. To determine the form of the
function $\overline{W}(x)$, we extend the natural integers $n$ to
complex values $s$ ($n\rightarrow s-1$) and we rewrite Eq. (\ref{5})
as
\begin{equation}
\int_0^\infty
x^{s-1}\overline{W}(x)\;dx=g(s-1,k)=\mathcal{M}\left[\overline{W}(x);s\right],
\end{equation}
on the other hand
\begin{equation}\label{h4}
\overline{W}(x)={1\over 2\pi
i}\int_{\mathcal{C}-i\infty}^{\mathcal{C}+i\infty}g(s-1)x^{-s}ds=\mathcal{M}^{-1}[g(s-1,k);x]
\end{equation}
where $\mathcal{M}\left[\overline{W}(x);s\right]$ is the Mellin
transform of $\overline{W}(x)$ and $\mathcal{M}^{-1}[g(s-1,k);x]$ is
the inverse Mellin transform of $g(s-1,k)$.

Finally, the Weight function may be written as
\begin{equation}\label{h5}
W(x)=\frac{\mathcal{M}^{-1}[g(s-1,k);x]}{\pi\mathcal{N}^2\left(x,k\right)},
\end{equation}
and Eq. (\ref{4}) holds.
 \medskip

Now, we are  going to build the coherent states of physical systems corresponding to particular
choices of the exponent $k$.

For $k=2$, the $GHA$ associated to the power-law potential spectrum
reduces to the Heisenberg algebra with characteristic function
$f(x,2)=x+1$. In this case, we have $N_{n-1}^2=n$ and Eq. (\ref{3})
becomes the standard coherent states for the harmonic oscillator,
$|z,2\rangle=\mathcal{N}\left(|z|^2,2\right)\sum_{n=0}^{\infty}
{z^n\over\sqrt{n!}}|n\rangle$, with normalization function
$\mathcal{N}\left(|z|^2,2\right)=\exp\left(-{|z|^2\over2}\right)$
and the Weight function ${1\over\pi}.$

In the limit case $k\rightarrow\infty$, the power-law potential
becomes the infinite square-well potential, and the energies defined
in Eq. (\ref{0}) are written as
\begin{equation}
E_{n,\infty}=(n+1)^2\qquad\qquad n=1,2,3,\cdot\cdot\cdot
\end{equation}
We can check that
\begin{equation}\label{c1}
E_{n+1,\infty}=\left(\sqrt{E_{n,\infty}}+1\right)^2.
\end{equation}
From Eq. (\ref{c1}) we find that the characteristic function for
this physical system is $f(x,\infty)=\left(\sqrt{x}+1\right)^2$  and
the corresponding $GHA$ generated by the Hamiltonian $H$ and the
creation and annihilation operators $A^\dag$, $A$ is given by
\begin{eqnarray}
\label{c2}\left[H,A^\dag\right]&=&2A^\dag\sqrt{H}+A^\dag, \\
\left[H,A^\dag\right]&=&-2\sqrt{H}A-A,\\
\label{c3} \left[A,A^\dag\right]&=&2\sqrt{H}+1.
\end{eqnarray}
Fock space representations of the algebra generated by $H$, $A$ and
$A^\dag$, as in Eqs. (\ref{c2})$-$(\ref{c3}), are obtained by
considering the eigenstates $|n\rangle$ of $H$. We can verify that
$N_{n-1}^2=(n+1)^2-1,$ and the action of this algebra generators on
$|n\rangle$ is given by
\begin{eqnarray}
H|n\rangle &=&\sqrt{n^2}|n-1\rangle, \qquad  n=1,2,\cdots,\\
A^\dag|n\rangle &=&\sqrt{n^2-1}|n+1\rangle,\\
 A|n\rangle &=&\sqrt{(n+1)^2-1}|n-1\rangle.
\end{eqnarray}

Therefore, the coherent states are
\begin{equation}
|z,\infty\rangle=\mathcal{N}\left(|z|^2,\infty\right)\sum_{n=0}^{\infty}
{z^n\over {N_{n-1}!}}|n\rangle
\end{equation}
where
\begin{equation}
N_{n-1}!={1\over\sqrt{2}}\sqrt{n!}\sqrt{(n+2)!}.
\end{equation}
The normalization condition requires that
\begin{equation}
2\mathcal{N}\left(|z|^2,\infty\right)\sum_{n=0}^{\infty}{|z|^{2n}\over
n!(n+2)!}=1.
\end{equation}
Noting that
\begin{equation}
\sum_{n=0}^{\infty}{|z|^{2n}\over n!(n+2)!}={I_2(2|z|)\over|z|^2},
\end{equation}
for $0\leq|z|<1$, where $I_n(z)$ is the modified Bessel function of
the first kind of order $n$. Thus, the normalization function my be
written as
\begin{equation}
\mathcal{N}\left(|z|^2,\infty\right)=\left({|z|^2\over2I_2(2|z|)}\right)^{1\over2}
\end{equation}
where $0\leq|z|<\infty$.

To resolve the identity operator in terms of the coherent states
corresponding to the square-well potential we need to find the
Weight function $W(x)$ satisfying
\begin{equation}\label{h5}
\int_0^\infty x^n\overline{W}(x)\;dx=n!(n+2)!, \qquad
 n=0,1,2,\cdot\cdot\cdot
\end{equation}
where $x=r^2$ and $\overline{W}(x)=\pi
W(x)\;\mathcal{N}^{\;2}\left(x,\infty\right)$.

From Eq. (\ref{h4}), the function $\overline{W}(x)$ becomes
\begin{equation}
\overline{W}(x)={1\over 2\pi
i}\int_{\mathcal{C}-i\infty}^{\mathcal{C}+i\infty}\Gamma(s)\Gamma(s+2)x^{-s}ds.
\end{equation}
For determining this function, we shall mainly use the following relations
satisfied by the Mellin transform
\begin{eqnarray}
M\left[x^bf\left(ax^h\right);s\right]&=&{1\over h}a^{-{s+b\over
h}}f^*\left({s+b\over h} \right),\qquad \{a,h\}>0,\\
 {1\over 2\pi
i}\int_{\mathcal{C}-i\infty}^{\mathcal{C}+i\infty}f^*(s)g^*(s)x^{-s}ds&=&\int_0^\infty
f\left(x\over t\right) g(t){dt\over t}
\end{eqnarray}
where $M[f(x);s]=f^*(s)$ and $M[g(x);s]=g^*(s).$

Employing the above relations and using  the modified Bessel
function of second kind $K_2(x)$\cite{29}, we find that
\begin{eqnarray}
\nonumber\overline{W}(x)&=&{1\over 2\pi
i}\int_{\mathcal{C}-i\infty}^{\mathcal{C}+i\infty}\Gamma(s)\Gamma(s+2)x^{-s}ds\\
\nonumber&=& \int_0^\infty t\exp\left(-t\right)\exp\left({-{x\over
t}}\right)\;dt\\
&=&2\;x\;K_2(2\sqrt{x}).
\end{eqnarray}
The Weight function $W(x)$ given in Eq. (\ref{h5}), allowing the
resolution of the identity operator, is written as
\begin{equation}
W(x)={2\;x\;K_2(2\sqrt{x})\over\pi\mathcal{N}^{\;2}\left(x,\infty\right)}.
\end{equation}
Finally,
\begin{equation}
W(x)={4\over\pi}K_2(2\sqrt{x})I_2(2\sqrt{x}).
\end{equation}

In order to complete the discussion, we should investigate the
statistical properties of the $GHA$ coherent states for power-law
potentials. To this end, we study Mandel's $Q$-parameter defined by
\cite{30}
\begin{equation}
Q={\langle(\Delta N)^2\rangle-\langle N\rangle\over \langle
N\rangle},
\end{equation}
where
\begin{equation}
\langle
N\rangle=\mathcal{N}^2\left(|z|^2,k\right)\sum_{n=0}^{\infty}n{|z|^{2n}\over
g(n,k)}
\end{equation}
is the average photon number of the state, and
\begin{equation}
\langle(\Delta N)^2\rangle=\langle N^2\rangle-\langle N\rangle^2.
\end{equation}

This parameter is a good indication to determine whether a state has
a sub-Poissonian (if $-1\leq Q<0$), Poissonian (if Q=0) or
super-Poissonian photon number distribution (if $Q>0$).

In Fig. 1,  we  plot the Mandel's $Q$-parameter in terms of the
$GHA$ coherent states amplitude $|z|$ for loosely binding potentials
($k<2$). From the figure we find that the states exhibit a
super-Poissonian behaviour for lower values of $|z|$ and
sub-Poissonian for high $|z|$. In the case of tightly binding
potentials ($k>2$), it is noticed that the Mandel's parameter is
always negative for all values of $|z|$ (see Fig. 2), so the
distribution is sub-Poissonian. However, we observe that the state
becomes in some sense {\it less classical} as $|z|$ becomes large.
By less classical we mean here that the states in question get
farther from the states exhibiting Poissonian statistics in
particular Glauber's coherent states.

It  is well known that the study of the statistical properties is an
important topic in quantum optics. In this way, our results show
that those coherent states may be useful to describe the states of
ideal and real lasers by a proper choice of the power-law exponent.



\section{Conclusion}
In this letter, using an algebraic approach, we constructed coherent
states for power-law potentials. The minimum set of conditions
required to obtain Klauder's coherent states is investigated. We
have shown that these states describe a large class of quantum
systems (harmonic oscillator, loosely binding potentials and tightly
binding), which can be used in several branches of quantum physics.

Finally, we investigated the statistical properties of these
coherent states using Mandel's $Q$ parameter. We have shown that
they exhibit Poissonian, sub-Poissonian or super-Poissonian
distributions. However, these states are useful to describe the
states of ideal and real lasers by a proper choice of the different
parameters (i.e., $z$ and $k$). These properties, are useful in
branches of quantum physics, such as quantum information. Indeed in
such studies, we are interested in the quantum behavior of these
states. So by a proper choice of the different parameters. One can
choose an adequate set of states in order to generate the
entanglement of bipartite composite systems using  a beam splitter.
This may open new perspectives to exploit these entangled states in
the context of quantum teleportation \cite{31}, dense coding
\cite{32} and entanglement swapping \cite{33}.

\section*{Acknowledgments}

 K.B. is supported by the National Centre for Scientific and Technical Researcher of Morocco. K.B.
 acknowledges the hospitality of the Abdus Salam International Centre for Theoretical Physics (Trieste, Italy).

\newpage

\begin{figure}
  \includegraphics[width=10cm]{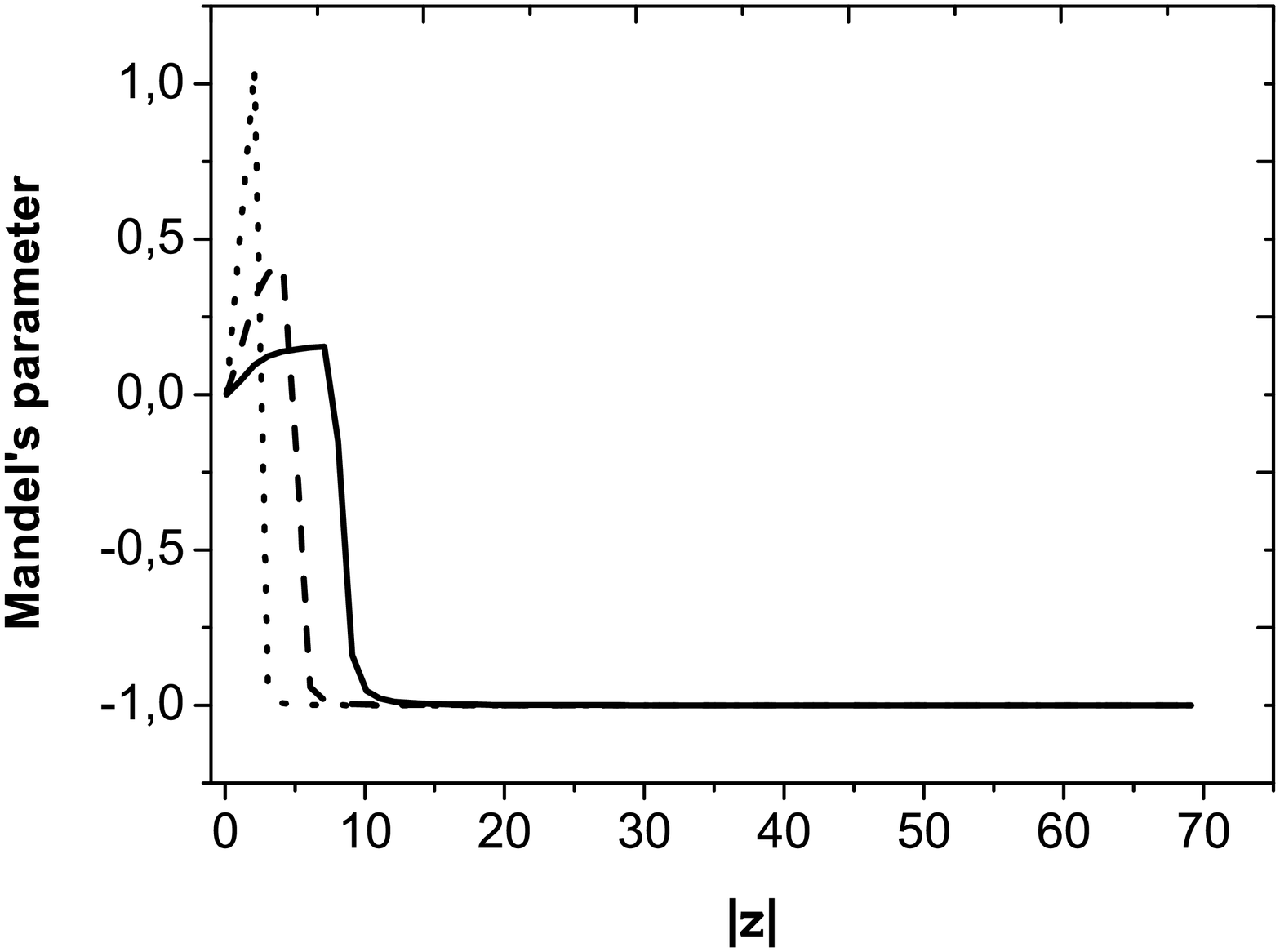}\\
  \caption{ Mandel's $Q$ parameter for loosely binding
potentials versus $|z|$  for $k=0.5$ (dotted line), $k=1$ (dashed
line) and $k=1.5$ (solid line).}\label{}
\end{figure}

\begin{figure}
  \includegraphics[width=10cm]{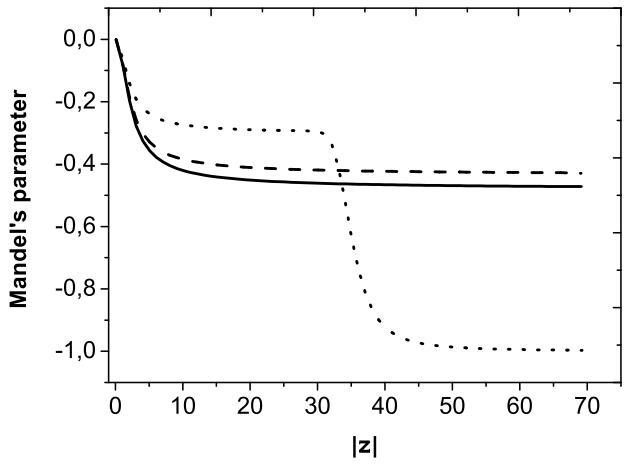}\\
  \caption{Mandel's $Q$ parameter for  tightly binding potentials versus $|z|$
 for $k=5$ (dotted line), $k=15$ (dashed line) and $k\rightarrow\infty$ (solid line).}\label{}
\end{figure}

\end{document}